%% file: paper.tex
\documentclass[sigconf]{acmart}

\usepackage{tabularx}
\usepackage{amsmath}
\usepackage{todonotes}
\usepackage{cleveref}
\usepackage{enumitem}
\usepackage{flushend}

\AtBeginDocument{%
  \providecommand\BibTeX{{%
    \normalfont B\kern-0.5em{\scshape i\kern-0.25em b}\kern-0.8em\TeX}}}

\setcopyright{acmcopyright}

\copyrightyear{2025}
\acmYear{2025}
\setcopyright{acmlicensed}\acmConference[SIGCSE 2025]{Proceedings of the 56th ACM Technical Symposium on Computer Science Education V. 1}{February 26 - March 01, 2025}{Pittsburgh, PA, USA}
\acmBooktitle{Proceedings of the 56th ACM Technical Symposium on Computer Science Education V. 1 (SIGCSE 2025), February 26 - March 01, 2025, Pittsburgh, PA, USA}
\acmPrice{15.00}
\acmDOI{}
\acmISBN{978-1-4503-XXXX-X/24/03}

\begin{document}

\title{CryptoEL: A Novel Experiential Learning Tool for Enhancing K-12 Cryptography Education}

\settopmatter{authorsperrow=2}
\author{Pranathi Rayavaram}
\email{nagapranathi_rayavaram@student.uml.edu}
\affiliation{%
  \institution{University of Massachusetts Lowell, Lowell, MA, USA}
  \country{}
}

\author{Ukaegbu Onyinyechukwu}
\email{Onyinyechukwu_Ukaegbu@student.uml.edu}
\affiliation{%
  \institution{University of Massachusetts Lowell, Lowell, MA, USA}
  \country{}
}

\author{Maryam Abbasalizadeh}
\email{maryam_abbasalizadeh@student.uml.edu}
\affiliation{%
  \institution{University of Massachusetts Lowell, Lowell, MA, USA}
  \country{}
}

\author{Krishnaa Vellamchetty}
\email{krishnaa_vellamchety@student.uml.edu}
\affiliation{%
  \institution{University of Massachusetts Lowell, Lowell, MA, USA}
  \country{}
}

\author{Sashank Narain}
\email{sashank_narain@uml.edu}
\affiliation{%
  \institution{University of Massachusetts Lowell, Lowell, MA, USA}
  \country{}
}

\renewcommand{\shortauthors}{Author 1 et al.}

\input{Abstract/abstract}


\maketitle

\input{Introduction/introduction}
\input{Related_Work/relatedwork}

\input{System_Design/systemdesign}
\input{Evaluation/evaluation}

\input{conclusion}

\bibliographystyle{ACM-Reference-Format}
\bibliography{paper}

\end{document}

%% file: Abstract/abstract.tex
\begin{abstract}
This paper presents an educational tool designed to enhance cryptography education for K-12 students, utilizing Kolb’s Experiential Learning (EL) model and engaging visual components. Our tool incorporates the four stages of EL—Concrete Experience, Reflective Observation, Abstract Conceptualization, and Active Experimentation—to teach key cryptographic concepts, including hashing, symmetric cryptography, and asymmetric cryptography. The learning experience is enriched with real-world simulations, customized AI-based conversation agents, video demonstrations, interactive scenarios, and a simplified Python coding terminal focused on cryptography. Targeted at beginners in cybersecurity, the tool encourages independent learning with minimal instructor involvement. An evaluation with 51 middle and high school students showed positive feedback from 93\% of participants, who found the simulations, visualizations, AI reflections, scenarios, and coding capabilities engaging and conducive to learning. Comprehension surveys indicated a high understanding of cryptography concepts: hashing (middle school: 89\%, high school: 92\%), symmetric cryptography (middle school: 93\%, high school: 97\%), and asymmetric cryptography (middle school: 91\%, high school: 94\%).
\end{abstract}

%% file: Introduction/introduction.tex
\section{introduction}

Cryptography is increasingly vital in K-12 education as students spend more of their lives online, necessitating an understanding of digital security. A 2023 Pew Research Center survey \cite{Pew2023} found that 95\% of U.S. teenagers have access to a smartphone, with nearly 46\% online constantly. Despite its importance, cryptography is absent from many K-12 education curricula. Fewer than 5\% of high school students and less than 1\% of middle school students in the U.S. receive formal education in cryptography \cite{cyberorg_state_2020}. Addressing this gap is crucial for equipping students with the skills to secure their digital lives and future careers.

The current educational approach to cryptography emphasizes advanced cybersecurity and mathematics \cite{educators-1,educator-2,educator-3}, leading to its exclusion from early education curricula. This is compounded by a lack of age-appropriate resources and engaging visual aids. To bridge this gap, structured educational frameworks are needed to simplify cryptographic principles for K-12 students, making them practical and engaging. Kolb’s Experiential Learning (EL) \cite{EL-book} addresses this through steps like concrete experience, reflective observation, abstract conceptualization, and active experimentation. Research shows that EL is highly effective for teaching complex concepts, enhancing retention, developing practical skills, and facilitating real-world application \cite{EL-1,EL-2,EL-3,EL-4}. By integrating theoretical knowledge with practical experiences, we can foster a deeper understanding of cybersecurity from an early age, building a strong foundation for future learning.

Despite initiatives like CryptoPals \cite{ncc_groups_cryptopals} and CryptoScratch \cite{cryptoscratch} expanding the K-12 cryptography curriculum with modern algorithms such as AES, RSA, and SHA, there is limited foundational material necessary for a thorough understanding of modern cryptography. Tools like Visual CryptoED \cite{visualcryptoed} address this gap through visualizations to explain cryptographic concepts but offer limited interactivity and hands-on coding experiences, which are crucial for grasping complex concepts~\cite{why-handson}. Research, such as Konak \cite{some-ini}, highlights the importance of experiential learning in K-12 cybersecurity education but focuses primarily on instructor-led sessions, leaving reflection and practical application to external activities. Therefore, there is a need for tools that integrate the full experiential learning cycle into cryptography education, encompassing real-world experiences, visualizations, reflection, and implementation, to provide a comprehensive and engaging learning experience.

\begin{figure}[t]
    \centering
    \includegraphics[width=1\linewidth]{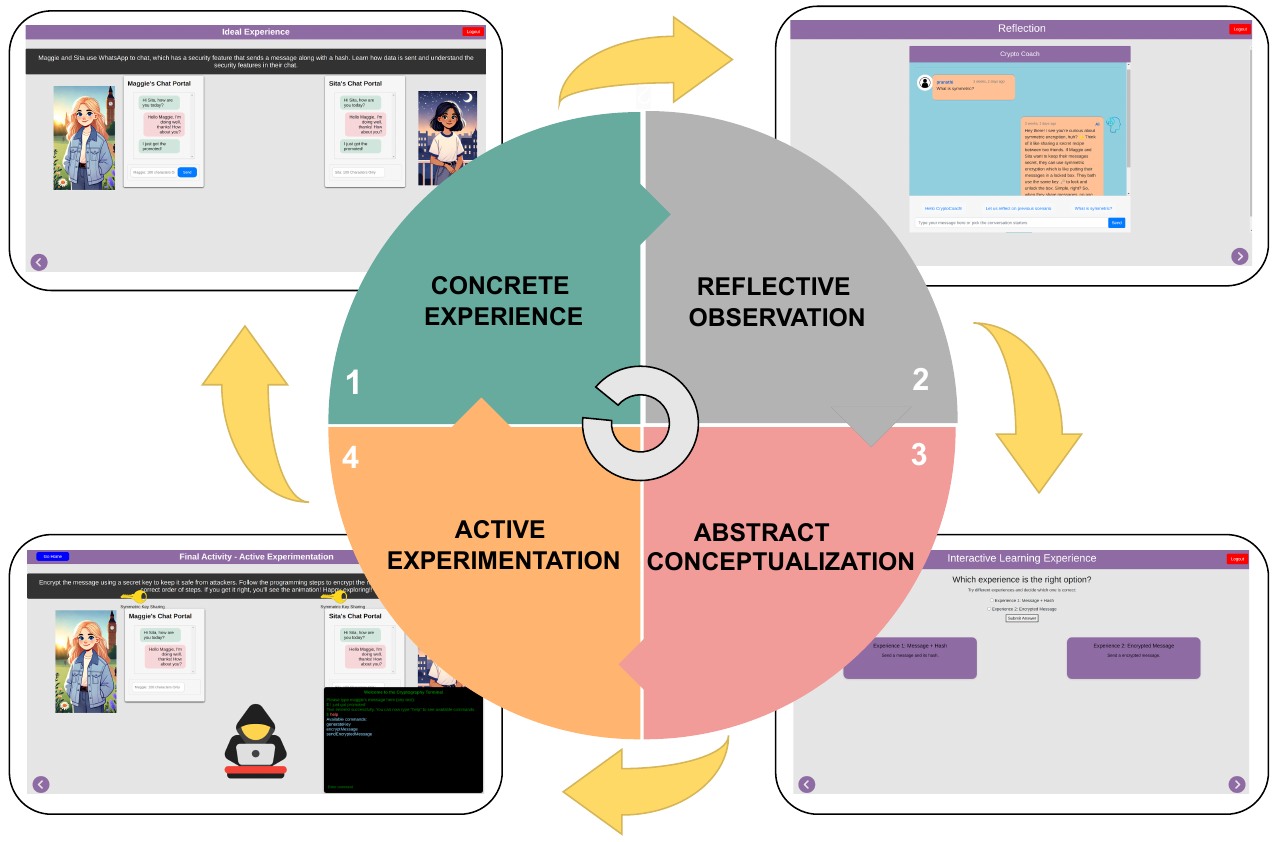}
    \caption{Integration of Experiential Learning Stages in Our Cryptography Education Tool.}
    \label{fig:image1}
    \vspace{-0.5em}
\end{figure}

This paper introduces a novel open-source tool \cite{Rayavaram_CryptoEL} for K-12 students that integrates the four stages of experiential learning (EL) to teach key cryptographic concepts: hashing, symmetric cryptography, and asymmetric cryptography. Our tool employs real-world visualizations of applications familiar to students, such as application portals, chat applications, and login interfaces, making learning relevant and engaging. In the concrete experience stage, students interact with real-world scenarios, experiencing both ideal conditions and how attackers can manipulate unsecured messages. During reflective observation, students analyze these experiences with customized AI-based conversation agents to understand why security incidents occurred. In abstract conceptualization, students interact with simulations, watch instructional videos, and learn cryptographic methods to prevent the observed attacks. Finally, in the active experimentation stage, students use a specialized terminal to execute simplified Python cryptography commands, reinforcing their understanding. This approach, illustrated in \Cref{fig:image1}, aims to ensure a clear, concise, and engaging learning experience that effectively teaches cryptography concepts.

A preliminary study involving 51 middle and high school students found the tool highly engaging and effective. Nearly 93\% of both middle and high school students found it very beneficial for their education. About 87\% of middle school and 79\% of high school students positively rated the hands-on activities. Students appreciated the engaging visualizations, AI chat reflections, and real-world examples, with comments such as, \textit{`The visualizations helped me because I am a visual learner'}, and \textit{`...This was the best tool by far and would greatly recommend it to others because it helps you understand the concept'}. Post-comprehension surveys, despite limited instructor involvement, showed high understanding: hashing (middle $\approx$89\%, high $\approx$92\%), symmetric cryptography (middle $\approx$93\%, high $\approx$97\%), and asymmetric cryptography (middle $\approx$91\%, high $\approx$94\%). These findings confirm the tool's effectiveness in teaching complex cryptographic concepts and promoting active learning through visual, engaging, and experiential methods.

In summary, this paper makes the following contributions:
\begin{itemize}[noitemsep,topsep=0pt]
\item We created an educational tool based on Kolb’s experiential learning model to teach cryptography using real-world applications: hashing with submission portals, symmetric cryptography via chat applications, and asymmetric cryptography through login interfaces. 
\item A study with 51 middle and high school students found that over 90\% of students showed a strong understanding of key concepts like hashing, symmetric and asymmetric cryptography. Students particularly appreciated the hands-on activities and real-world applications.
\end{itemize}

The rest of this paper is structured as follows: Section 2 reviews related work. Section 3 details the design and implementation of our educational tool. Section 4 presents the user study, evaluates the tool's effectiveness, and discusses student experiences. Section 5 concludes the research.

\begin{figure*}[t]
    \centering
    \includegraphics[width=0.99\textwidth, height=0.13\textheight]{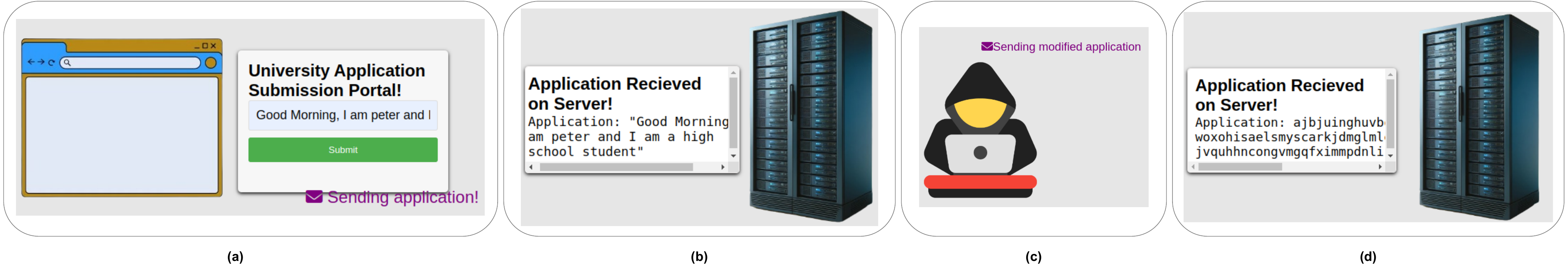}
      \caption{Ideal and Attacker Experiences in the Hashing module.}
    \label{fig:Experience}
\end{figure*}

%% file: Related_Work/relatedwork.tex
\section{Related Work}\label{sec:relatedwork}
Many cryptographic tools \cite{cyberchef,crypto-tutor,cryptomaster,cryptolabs,CryptoTool,nathan} target advanced audiences with backgrounds in mathematics and programming, making them unsuitable for K-12 education. In contrast, K-12 tools often focus on basic techniques like Caesar ciphers \cite{bassi2023cybersecurity,sudha2023changes} with limited initiatives supporting modern cryptographic algorithms and protocols. These initiatives offer modern cryptographic learning through challenges \cite{ncc_groups_cryptopals}, workshops \cite{gencyber}, unplugged activities \cite{feaster_teaching_2011,konak2014}, and games \cite{riskio,YAMIN2021102450}, but typically cover specific algorithms without comprehensive integration. Therefore, we will analyze works that integrate cryptography into K-12 contexts using modern algorithms and protocols, and are adaptable to more advanced concepts.

Efforts to incorporate modern cryptography into K-12 education include block-based programming tools like Lodi et al.'s Snap! which focuses on encryption and decryption \cite{lodi_cryptography_2022}. CryptoScratch \cite{cryptoscratch}, a Scratch extension, allows students to experiment with hashing, symmetric cryptography, and asymmetric cryptography. However, these tools primarily enable students to implement modern cryptography systems and do not provide comprehensive educational materials. To address this gap, works such as Rayavaram et al. \cite{rayavaram} and Visual CryptoEd \cite{visualcryptoed} have developed interactive, visual, and role-playing modules to teach concepts like integrity protection, confidentiality, and authentication using hashing, symmetric cryptography, and asymmetric cryptography. While these approaches make learning more engaging, they primarily focus on initial exposure to concepts and offer limited opportunities for students to reflect on their learning or implement cryptographic techniques in practical scenarios. Some initiatives \cite{K12-SEL,K12-SEL2} have integrated research-based models such as Kolb’s Experiential Learning (EL) into cryptography education. The most relevant to our work is Konak \cite{some-ini}, which focused on the Active Experimentation stage of EL using the tool CrypTool \cite{CryptoTool}, but left the crucial phases of Concrete Experience, Reflection, and Abstract Conceptualization as external activities guided by instructors.

Our tool stands out by fully integrating all four stages of Kolb’s EL cycle, enhancing learning through interactive elements that personalize the experience for each student. It includes visualizations and interactive attacker simulations that students can experience to understand real-world attacks on realistic applications. Students reflect on these attacks with AI-enabled conversations, view videos and mitigation scenarios to deepen abstract conceptualization, and engage in active experimentation through an embedded Python cryptography terminal. These features foster a deeper understanding of cryptographic concepts and equip students with practical programming skills, providing a well-rounded and engaging educational experience in cryptography.

%% file: System_Design/systemdesign.tex
\section{Tool Design and Implementation}\label{sec:design}

This section provides an overview of the design and implementation of our tool, which is fully open-sourced \cite{Rayavaram_CryptoEL} to promote community collaboration and integration into K-12 classrooms.

\begin{figure*}[t]
    \centering
    \begin{minipage}{0.34\textwidth}
        \centering
        \includegraphics[width=.84\textwidth]{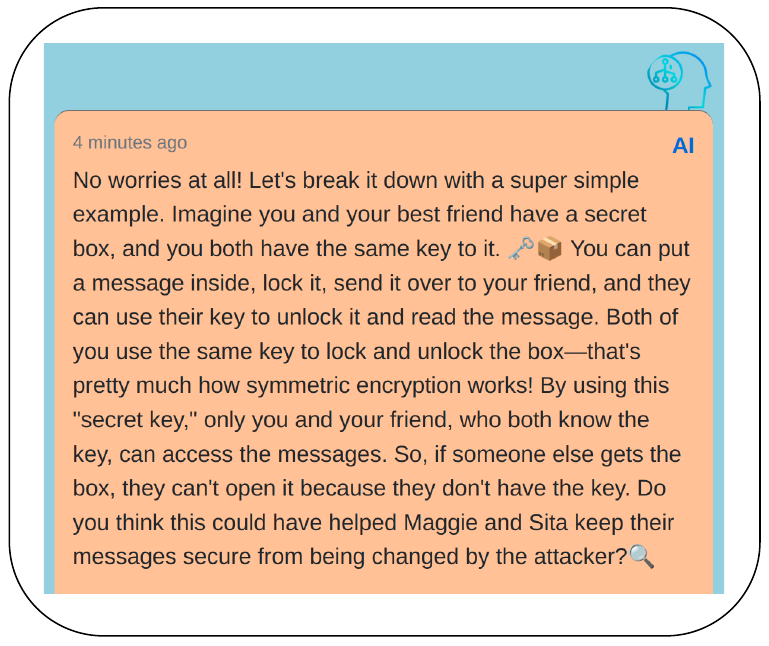}
        \caption{CryptoCoach Reflection Example.}
        \label{fig: Reflection}
    \end{minipage}
    \hfill
    \begin{minipage}{0.6550\textwidth}
        \centering
        \includegraphics[width=.99\textwidth]{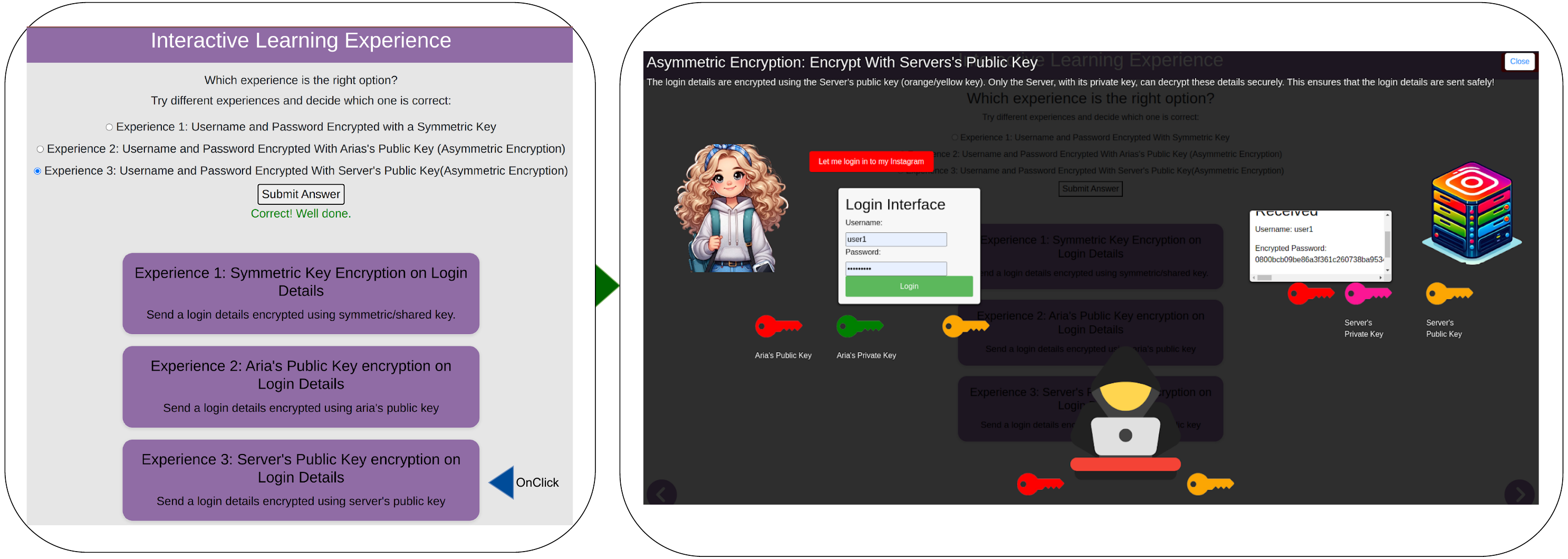}
        \caption{Abstract Conceptualization in Symmetric Cryptography module.}
        \label{fig:image3}
    \end{minipage}
\end{figure*}

\subsection{Overview}

CryptoEL is a web-based tool designed to teach complex cryptography concepts through Kolb's Experiential Learning (EL) model's four stages.  We chose the EL model for its emphasis on learning through direct experience, integrating theoretical concepts with real-world applications—crucial for cybersecurity education. Each concept—hashing, symmetric cryptography, and asymmetric cryptography—builds on the previous one, reinforcing and expanding the learner's understanding. We used the scalable and robust Django web framework \cite{django} as the backend to manage the application's functionality and data flow. Real-world scenarios, crafted with HTML, CSS, and JavaScript, make learning relevant and engaging. For instance, we use a submission portal to teach hashing, a chat interface for symmetric cryptography, and a login interface for asymmetric cryptography. This approach helps students relate to the material and apply cryptographic concepts in practical contexts. Additionally, all text in the tool has been checked for age-appropriateness using the Kincaid calculator \cite{kincaid1975derivation}.

In Kolb's Experiential Learning model, the goal of the concrete experience stage is to immerse users in engaging scenarios that provide hands-on learning opportunities. We incorporated this principle in our tool through two interactive visual simulations: an Ideal Experience demonstrating how a message would ideally be transmitted over a network, and an Attack Experience, where the same message is intercepted and modified by an attacker, highlighting specific vulnerabilities. An example of such an experience is illustrated in \Cref{fig:Experience} and further detailed in \Cref{sec:cryptoconcepts}. By contrasting these scenarios, we help students understand how vulnerabilities can be exploited, motivating them to reflect on what went wrong and how to address it. To ensure the effectiveness of these immersive experiences, we use visualizations and animations that clarify complex ideas and transform abstract concepts into engaging, accessible learning opportunities.

In the second stage of Kolb's Experiential Learning, reflection helps users analyze and understand their previous concrete experiences. We have integrated this reflective process into our tool using customized ChatGPT models from OpenAI, called "CryptoCoach." CryptoCoach features three models specifically tailored to hashing, symmetric cryptography, and asymmetric cryptography. It engages students by asking questions that encourage reflection on their experiences, emphasizing the implications of weak cryptography. A good illustration of CryptoCoach is shown in \Cref{fig: Reflection}, where it provides a simple overview of symmetric encryption and then asks another question to prompt further reflection. We have trained and extensively tested CryptoCoach to ensure it effectively supports learning and aligns with the educational objectives of each topic. Our goal is to provide a valuable and accessible learning experience while adhering to ethical AI usage standards.

In the third stage of Kolb's Experiential Learning model, abstract conceptualization helps students develop a structured understanding of their learning. We incorporated this stage into our tool using two distinct components. First, students watch an illustrative video introducing the most appropriate cryptography concept to mitigate the attack they experienced in the first stage. Following the video, students can view multiple simulations: some remain vulnerable to attacks, while one demonstrates adequate protection using the concept from the video. As shown in \Cref{fig:image3}, students are presented with three choices. Selecting any option runs the specific scenario with visualizations and animations similar to the experience stage. Further details are provided in \Cref{sec:cryptoconcepts}. By interacting with these scenarios, students can explore different strategies and understand why the correct solution is effective.

Finally, in Kolb's Experiential Learning model, active experimentation involves applying new knowledge through hands-on activities. We incorporated this into our tool with a final activity where students use an embedded terminal to practice cryptographic commands for secure data transmission. This terminal features our custom simplified commands, enabling students to easily implement the secure abstract conceptualization scenario from the previous stage. These commands provide comprehensive feedback, helping students determine the correct ordering of cryptographic algorithms for secure data transfer. Successful execution generates an animation that illustrates this data transfer, reinforcing the correct procedures. The commands internally call functions from the Python library pycryptodome \cite{pycryptodome} to leverage robust and efficient existing open-source implementations.
 
\subsection{Tool Implementation}\label{sec:cryptoconcepts}

This section introduces the key cryptography concepts in our tool: hashing, symmetric cryptography, and asymmetric cryptography. Building on Rayavaram et al.'s work~\cite{visualcryptoed}, we focus on these concepts to establish robust comprehension of basic cryptographic principles. More advanced topics, such as Pretty Good Privacy (PGP) \cite{pgp}, Digital Signatures \cite{digitalsignatures}, HMAC \cite{hmac}, and Transport Layer Security (TLS) \cite{python-ssl}, will be covered in future work to ensure a thorough and incremental learning process.

\vspace{0.25em}
\noindent\textbf{Hashing:} This module highlights the importance of data integrity and the consequences of tampering through a university application portal scenario, teaching students the critical role of hashing. Students begin the experience stage by watching a video of Peter, a high school student, submitting his application. In the ideal scenario, they type the application (\( M \)) and see it sent successfully to the university server, demonstrating normal operations. The scenario then shifts to an attacker modifying the text to \( \hat{M} \), which the server accepts without verification, exposing vulnerabilities (\Cref{fig:Experience}). During the reflection stage, students interact with CryptoCoach, our AI assistant, who discusses the attack and introduces hashing as a mitigation solution. In the abstract conceptualization stage, students watch a video explaining hashing fundamentals and then engage with three scenarios to identify the correct solution: (1) Sending the plain text \( M \), which the attacker modifies to \( \hat{M} \) (insecure), (2) Sending only the hash \( H(M) \), which the server cannot parse (incorrect), (3) Sending both text and hash values \( (M, H(M)) \), which the server correctly handles (secure). Here, \( H(.) \) represents a secure hashing function like SHA-256. In the secure scenario (3), students see the portal verifying the message by recomputing the hash, comparing it with the original hash, and accepting the message if they match. While hashing alone is not sufficient for complete security, it is presented as the best option in this context to help students understand the concept. This concept is later expanded in the symmetric cryptography module to enhance security further. After interacting with each option, students understand that sending both text and hash \( (M, H(M)) \) is the best approach to detect tampered messages and maintain data integrity. In the final activity, students execute the commands `generateHash' and `sendMessageHash' in the embedded terminal, allowing them to implement the correct solution on their own and reinforcing their understanding through hands-on practice.

\vspace{0.25em}
\noindent\textbf{Symmetric Cryptography:} This module builds on the foundational concepts of hashing by introducing symmetric cryptography through a chat interface, helping students understand data confidentiality and encryption mechanisms for secure communications. It begins with a video of best friends Mary and Sita discussing Mary's promotion. The student assumes Mary's role and sends a message to Sita. Ideally, when a student types a message (\( M \)), the message and its hash value \(M^H = (M, H(M))\) are transmitted to Sita. In the attacker scenario, students see that an attacker modifies both the message and the hash value \(\hat{M}^H = (\hat{M}, H(\hat{M}))\), which is falsely accepted by Sita's chat portal as, ``I lost my job, can you transfer money to my account.'' This observation becomes the topic of reflection with CryptoCoach, prompting students to analyze the attack and explore symmetric cryptography as a solution (\Cref{fig: Reflection}). Moving to abstract conceptualization, a follow-up video explains symmetric cryptography fundamentals, including key generation and encryption, and provides three interactive simulations to identify the correct secure solution: (1) Sending the text message and hash value \(M^H\), which the attacker changes to \(\hat{M}^H\) (insecure), (2) Encrypting the message and hash, \( C=E(K,M^H) \), with a pre-shared secret key \(K\), which an attacker can no longer read or modify (secure), and (3) Encrypting the message with its hash value \(E(H(M), M)\), which the receiver portal is unable to parse (incorrect). Here, \(E(K, M)\) represents message \(M\) encrypted using key \(K\) with a secure algorithm such as AES. Students see that, in scenario (2), Sita's chat portal can decrypt the message \(M^H\) using the decryption function \(D(K, C)\), verify the hash, and accept the message, demonstrating that symmetric cryptography prevents the attacker from reading the message and preserves confidentiality. In the final experimentation stage, students use our embedded terminal to execute commands: `generateKey,' `encryptMessage,' and `sendEncryptedMessage,' to create the secure solution they observed. Feedback, such as ``There is no key for encryption. Please generate a key to perform the encryption,'' ensures proper command sequencing and reinforces the learning experience.

\vspace{0.25em}
\noindent\textbf{Asymmetric Encryption:} The asymmetric cryptography module uses a social media login interface to help students understand public and private keys, essential for protocols like TLS to ensure digital communication confidentiality and authenticity. It begins with a video featuring Aria, a high school student logging into social media from an unsecured network. After the video, students engage in a simulation of an ideal login experience. When they enter Aria's username and password \(M = (U, P)\), an animation shows the credentials being encrypted with a shared secret key \(K\), i.e., \(C=E(K,M)\), and safely transmitted to the server for decryption. Students then experience an attacker scenario, highlighting the vulnerability of symmetric cryptography, which requires pre-sharing a secret key \(K\). They observe that Aria's app and the server share the key \(K\) over an insecure channel, and it gets intercepted by the attacker, who then decrypts the credentials, i.e., \(M=D(K,C)\). In the reflection stage, students interact with CryptoCoach, who discusses the limitations of symmetric cryptography and introduces asymmetric cryptography, which doesn't require pre-sharing a secret key. The abstract conceptualization phase follows with a video explaining asymmetric cryptography fundamentals, including generating and using public and private keys for both Aria and the server. Although best practices in cryptography involve keys signed by a trusted CA, this introduction simplifies the concept. Students then see interactive simulations where Aria and the server generate their keys and share their public keys. They are presented with three options for encrypting credentials to determine the most secure method: 1) Using a symmetric key for encryption, i.e., \(C=E(K,M)\), where an attacker steals the credentials (insecure); 2) Encrypting the credentials with Aria's public key, i.e., \(\mathcal{E}(A_{pub}, M)\), which the server cannot decrypt without Aria's private key \(A_{pri}\) (incorrect); 3) Encrypting the credentials with the server's public key, i.e., \(\mathcal{E}(S_{pub}, M)\), which only the server can decrypt using its private key \(S_{pri}\) (secure). Here, \(\mathcal{E(.,.)}\) denotes asymmetric encryption using a secure algorithm like RSA. Through interacting with these scenarios, students not only conceptualize the benefits of not sharing a secret key but also understand the importance of safeguarding the private key in asymmetric cryptography. In the final active experimentation activity, students apply their understanding through a hands-on terminal exercise shown in \Cref{fig:image4}. Typing `help' provides instructions for commands such as `generateAriaPrivateKey,' `generateAriaPublicKey,' `grabServerPublicKey,' `encryptMessageServerPublicKey,' and `sendEncryptedMessage.' Guided by this feedback, students learn to implement a secure login module using asymmetric cryptography.

%% file: Evaluation/evaluation.tex
\section{Evaluation}\label{sec:evaluation}

In this section, we discuss the user study design and evaluate our tool's effectiveness in teaching cryptographic concepts to students using pre- and post-surveys and concept quizzes.

\subsection{User Study}\label{sec:userstudy}

\vspace{0.25em}
\noindent\textbf{Participant Recruitment and Demographics:} We recruited 51 middle and high school students using purposive sampling after receiving IRB approval. Flyers were distributed through academic departments, student organizations, and personal contacts. Parental consent forms were obtained for minors. The online workshop included 18 middle school students (5 females, 13 males) and 33 high school students (15 females, 18 males) from diverse locations, primarily Texas, Virginia, and other US states. Internationally, a count of 7 students were from Kuwait, India, Canada, and Spain.

\begin{figure}[t]
\centering
\includegraphics[width=0.70\linewidth]{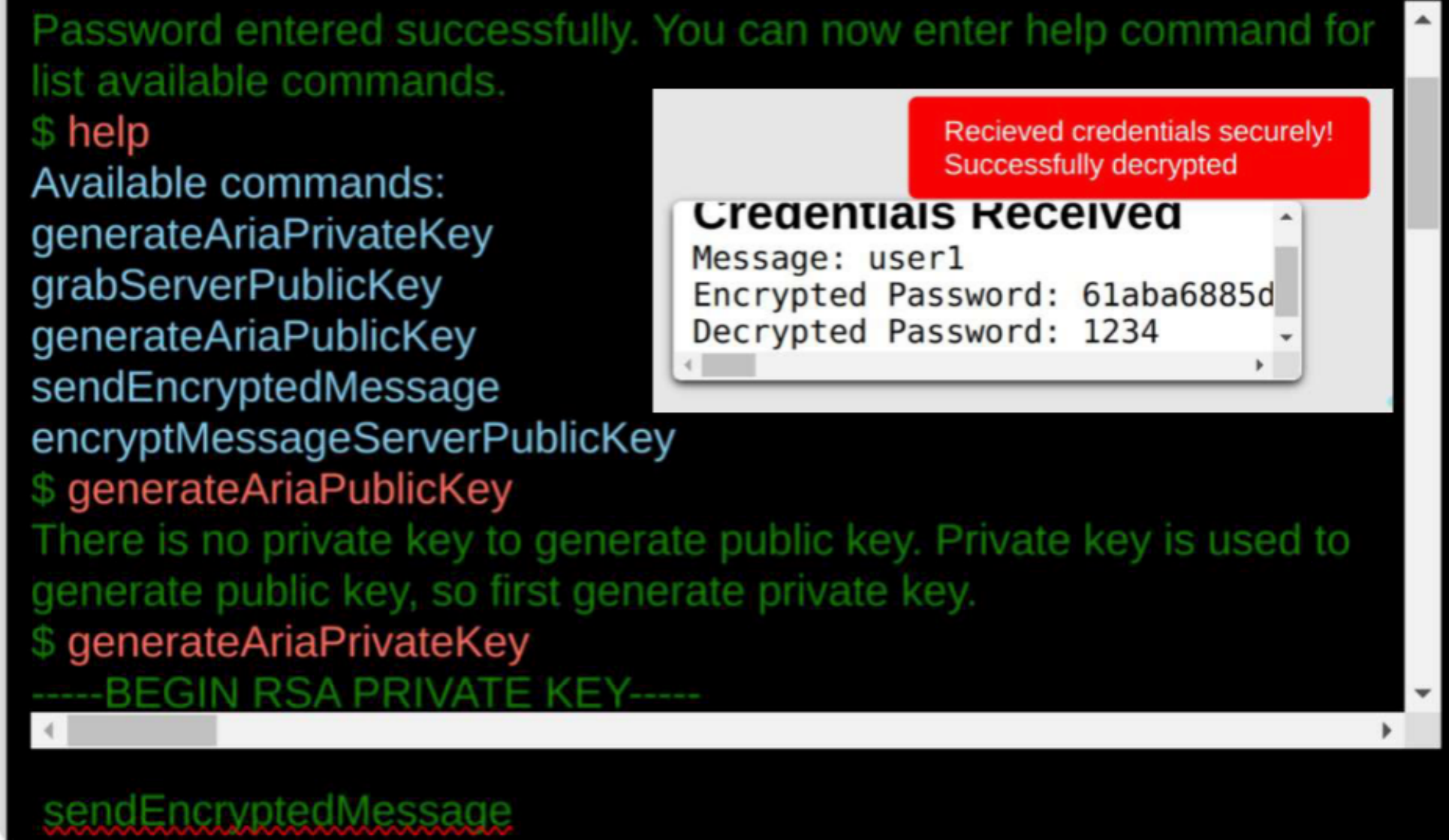}
    \caption{Asymmetric Cryptography Active Experimentation.}
    \label{fig:image4}
\end{figure}

\vspace{0.25em}
\noindent\textbf{Survey Design:} We conducted a pre-survey with multiple-choice questions to assess students' foundational knowledge in cryptography. As the workshop progressed, we introduced key cryptographic concepts (hashing, symmetric cryptography, asymmetric cryptography), followed by concept quizzes assessing comprehension of these topics. A post-survey, including the same questions as the pre-survey, was conducted to measure knowledge gains and evaluate the tool's effectiveness. Additional questions assessed students' perceptions, confidence, and overall satisfaction.

\vspace{0.25em}
\noindent\textbf{Study Design and Execution}: A 4-hour workshop was conducted to assess the tool's effectiveness. Students were assigned pre-created user credentials one day before the workshop. At the workshop's beginning, they completed the pre-survey questionnaire. After the pre-survey, we provided an overview of the user interface and navigational components. Students then interacted with the first two stages on their own, experiencing the ideal and attacker scenarios, and reflecting on these using CryptoCoach. This was followed by a 5-10 minute instructor-led discussion where students shared insights from their experiences and AI interactions. Instructors clarified concepts and addressed questions. Students then proceeded through the next two stages of abstract conceptualization and active experimentation on their own. At the end of the module, a 5-minute discussion ensured clarity and addressed any remaining doubts, followed by a conceptual quiz. This approach was replicated across all modules. After all modules were completed, students filled out a post-survey questionnaire.

\subsection{Comprehension of Cryptography Concepts}\label{sec: concepts_eval}

This section highlights students' comprehension of cryptography concepts taught using our tool. After each cryptography module, we assessed their understanding by distributing ten multiple-choice questions across three categories. The conceptual understanding category included five theoretical questions to evaluate students' grasp of the concepts. For example, ``What are the steps and order to achieve asymmetric encryption?'' assessed understanding of the public key encryption process. The practical application category had three questions to assess students' ability to apply the concepts in different scenarios. For example, ``We have a digital library with important old documents. How can hashing help keep these documents real and unchanged?'' assessed understanding of using hashing to ensure data integrity. The security implications category posed two questions to evaluate their understanding of how the concept contributes to security. For example, ``What happens if someone steals the symmetric key during transmission?'' tested knowledge of potential vulnerabilities and security measures.

\vspace{0.25em}
\noindent\textbf{Hashing Comprehension:} We assessed students' understanding of hashing through questions on techniques, command sequences, data storage applications, and password protection. The results showed high comprehension: 89\% for middle school students (conceptual: 85\%, practical: 89.58\%, security: 93.75\%) and 92\% for high school students (conceptual: 95\%, practical: 94.79\%, security: 87.50\%). These results suggest the tool effectively teaches hashing concepts across different age groups. High school students excelled in conceptual and practical comprehension, likely due to greater experience. Middle school students had a firmer grasp of security implications, which may be attributed to the tool’s emphasis on real-world scenarios that resonate well with their learning stage. 

\vspace{0.25em}
\noindent\textbf{Symmetric Cryptography Comprehension:} We assessed students' comprehension of symmetric cryptography concepts by evaluating its operational efficiency in encrypting large data, the importance of key protection, practical applications in securing online transactions, and the security risks with key management. Overall, 93\% of middle school students (conceptual: 82.35\%, practical: 96.08\%, security: 100\%) and 97\% of high school students (conceptual: 93.15\%, practical: 100\%, security: 98.44\%) demonstrated a strong grasp of symmetric encryption principles and key confidentiality. Both groups showed a better understanding of symmetric cryptography than hashing, likely because the concept of secrets, such as passwords, is more intuitive and directly applicable to real-life scenarios they are familiar with.

\vspace{0.25em}
\noindent\textbf{Asymmetric Cryptography Comprehension:} We assessed students' grasp of asymmetric cryptography by highlighting its key differences from symmetric cryptography, including the benefits of separate keys for encryption and decryption. The assessment covered practical applications for securing communications and security implications, emphasizing the importance of correctly sharing the public key and securing the private key. Overall, 91\% of middle school students (conceptual: 95.00\%, practical: 95.83\%, security: 81.25\%) and 94\% of high school students (conceptual: 99.33\%, practical: 96.67\%, security: 86.67\%) answered correctly. These findings suggest that students easily understood asymmetric cryptography and its practical use with our tool. However, both groups had lower comprehension of security implications due to the complexity of managing and safeguarding private keys.
		
\subsection{Pre- and Post-survey Comparison}\label{sec:userstudy}

We conducted paired t-tests to evaluate changes in middle and high school students' understanding of concepts before and after using our tool. Pre-surveys assessed students' familiarity with hashing, symmetric, and asymmetric cryptography, while post-surveys gauged their confidence and perceived knowledge improvement, using a Likert scale. The analysis revealed significant increases in understanding across all concepts. For middle school students, the t-statistics and p-values were: Hashing (t = 10.46, p = 4.72×10$^{-7}$), Symmetric Encryption (t = 8.85, p = 2.47×10$^{-6}$), and Asymmetric Encryption (t = 6.17, p = 7.04×10$^{-5}$). For high school students, the results were even more pronounced: Hashing (t = 20.76, p = 2.13×10$^{-16}$), Symmetric Encryption (t = 21.17, p = 1.40×10$^{-16}$), and Asymmetric Encryption (t = 27.13, p = 5.75×10$^{-19}$). These findings, with p-values well below 0.05, indicate a low likelihood that the observed improvements are due to chance. The significant increase among both middle and high school students highlights the tool's effectiveness in enhancing cryptography education.

\subsection{Student Satisfaction and Experiences}\label{sec:userstudy}

The CryptoEL tool was highly rated for engagement, effectiveness, and educational benefit, with nearly 93\% of both middle and high school students rating it 4 or above on a 5-point Likert scale. About 87\% of middle school students and 79\% of high school students gave positive feedback on the tool's hands-on nature. The active experimentation stage, where students implemented their own secure solutions, received high ratings from about 93\% of middle school and 89\% of high school students. About 60\% of middle school students and 89\% of high school students said they benefited from the scenarios in the experience and abstract conceptualization stages. For the AI-driven reflection stage, the numbers were 73\% of middle and 86\% of high school students. Next, we focus on specific student experiences, feedback, and instructor observations, further exploring the impact and success of the tool.

Middle school students were particularly engaged by the conversational agent's tailored responses, visualizations, and straightforward scenarios. One student said, ``I liked the AI chat reflections and how we were able to complete the activities hands-on,'' while another noted, ``The visualizations helped me because I am a visual learner, so when I was answering the questions, I was able to use the visualization.'' High school students engaged deeply with the practical applications of secure communication, using commands in the terminal window. One student commented, ``I liked how the commands used by the terminal window visualized how the server encrypts messages. This was the best tool by far and I would greatly recommend it to others because it helps you understand the concept.'' Their feedback highlighted the realistic use of commands and the terminal's role in demonstrating secure processes. High school students also demonstrated heightened engagement and curiosity. One remarked, ``I really liked how it got you physically engaged with all the concepts in cybersecurity. It was fun and motivated my curiosity for the topic. I would definitely recommend this tool to others trying to learn the same concepts.'' They also asked thoughtful questions such as, ``Can hash value be reversed to text?'' demonstrating that they were actively engaging with and critically thinking about the material. In contrast, middle school students relied more on analogies and visual supports, generally not extending their inquiries beyond the provided material.

However, some areas for improvement were noted. In particular, some students found the animations too fast and the AI responses lengthy. Addressing these issues by adjusting the activity pace, including analogies, simplifying navigation, and ensuring concise AI responses could further enhance the overall learning experience.

%% file: conclusion.tex
\section{Conclusion}\label{sec:conclusion}


This paper introduced a K-12 cryptography tool based on Kolb’s Experiential Learning, using simulations, AI reflections, and coding to teach core concepts. Evaluation shows high engagement and comprehension, addressing gaps in early cybersecurity education.
